\documentclass[12pt]{article}
\usepackage{epsfig}

\def\beq{\begin{equation}}
\def\eeq#1{\label{#1}\end{equation}}
\def\eeqn{\end{equation}}

\def\beqa{\begin{eqnarray}}
\def\eeqa#1{\label{#1}\end{eqnarray}}
\def\eeqan{\end{eqnarray}}

\let\bar=\overbar

\def\etal{{\it et al.}}

\def\D{{\cal D}}

\def\Dslash{\not{\hbox{\kern-4pt $D$}}}
\def\dslash{\not{\hbox{\kern-2pt $\del$}}}

\def\msb{{\bar{\ssstyle M \kern -1pt S}}}

\def\Title#1{\begin{center} {\Large {\bf #1} } \end{center}}
\def\D0{D0}
\begin{document}

\begin{raggedleft}
IUHEX-201 \\
\end{raggedleft}

\Title{Status of \D0 for $B$ Physics}

\bigskip\bigskip


\begin{raggedright}  

{\it R. Van Kooten \index{Van Kooten, R.} for the \D0 Collaboration\\
Physics Department\\
Indiana University\\
Bloomington, IN 47405\\
rickv@fnal.gov}
\bigskip\bigskip
\end{raggedright}

\section{Introduction}

In light of the voluminous data being collected at the $B$ factories,
it is worthwhile examining why it is interesting to study $B$ physics
at the Tevatron at Fermilab (Batavia, Illinois).  
Both \D0 and CDF enjoy a large rate of production
with $\sigma(p \bar{p} \rightarrow b \bar{b}) \approx 150$~$\mu$b
at a collision energy of 2 TeV compared to
$\sigma(e^+ e^- \rightarrow b \bar{b}) \approx 7$~nb at the
$Z^0$ peak, and 
$\sigma(e^+ e^- \rightarrow B \bar{B}) \approx 1$~nb at the
$\Upsilon(4S)$ peak.  
Also, in contrast to the $B$ factories
running at the $\Upsilon(4S)$ peak where only $B^0_d$ and 
$B^{\pm}$ are produced, all $b$ hadron species including
$B_s$, $B_c$, and $\Lambda_b$ are produced.
Measurements of $B_s$ mixing are important in the understanding of
the CKM triangle, and a great deal can be learned from the properties
and behavior
of hadrons containing heavier quarks besides the $b$.

\section{Run 2 $B$ Physics}

Run 1 for \D0 ended in 1996 and work began on the Main Injector and 
upgrading the detector as described in the next section.  
Run 2 is defined by the running of the 
Tevatron with the Main Injector at an increased collision energy 
of 2~TeV (from 1.8~TeV) and increased luminosity.  Run 2a is planned
to have an integrated luminosity of 2~fb$^{-1}$ and Run 2b of 
15--20~fb$^{-1}$.

Some of the more important topics of the planned \D0 Run 2
$B$ physics program are summarized in Table~\ref{tab:physics}.
Question marks following the decays $B \rightarrow \pi^+ \pi^-$ and 
$B_s \rightarrow K^+ K^-$ imply that although effort will be made to 
isolate these samples, it is unclear whether \D0's triggering capability
will allow this collection with reasonable bandwidth.

\begin{table}[htb]
\begin{center}
\begin{tabular}{l|l}
\hline
QCD tests & cross sections, correlations, \\
             charmonium polarization \\
CP violation and CKM angles & $\sin{2\beta}$ through 
$B \rightarrow J/\psi K_S$; \\
      & $\alpha$, $\gamma$ through 
$B \rightarrow \pi^+ \pi^-$?, $B_s \rightarrow K^+ K^-$? \\ 
Non-SM CP Violation & $B_s \rightarrow J/\psi \phi$ \\
$B_s$ mixing &  $B_s \rightarrow D_s n\pi$, 
                $B_s \rightarrow D_s \ell \mu$ \\
Spectroscopy and Lifetimes & $B^0_d$, $B^+$, $B^0_s$, $B_c$, $\Lambda_b$, 
double heavy baryons \\
Rare decays & $B \rightarrow \ell^+ \ell^- X_s$, 
              $B \rightarrow \ell^+ \ell^-$ \\ \hline
\end{tabular}
\caption{Summary of important topics in planned $B$ physics program.}
\label{tab:physics}
\end{center}
\end{table}

\section{Upgraded \D0 Detector}

The described program of $B$ physics could only be undertaken 
if the Run 1 \D0 detector were upgraded.  
The Run 2 detector
upgrade
retained the excellent Run~1 liquid argon/uranium calorimeter, 
while increasing the speed
of its readout. The muon toroids were retained, and the muon
system was upgraded for better muon 
identification and triggering, particularly suited for tagging
$b$'s decaying semileptonically.  
Most importantly for $B$ physics, a new tracker
operating in a solenoidal magnetic field was added.
Further details of the detector can be found elsewhere~\cite{detect}.

\begin{figure}[htb]
\begin{center}
\epsfig{file=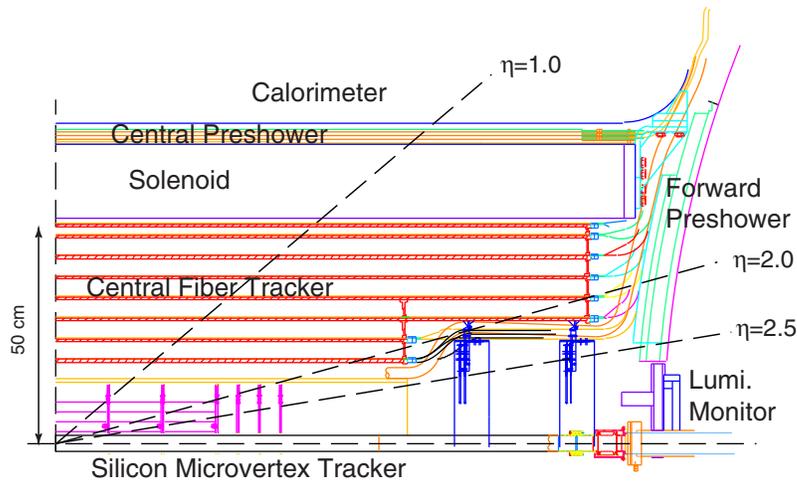,height=2.5in}
\caption{\D0 Upgrade detector tracking system.}
\label{fig:tracker}
\end{center}
\end{figure}

Figure~\ref{fig:tracker} illustrates the new tracker.
Working outwards in radius, the silicon microvertex detector (SMT)
consists of six barrels (four layers) of single and double-sided
silicon strip sensors interspersed with double-sided disks
providing 800k channels with a single hit space resolution
of 10~$\mu$m, leading to good impact parameter resolution.  
The central fiber tracker (CFT)
provides the necessary momentum resolution, and is made up 
of eight concentric barrels of scintillating optical fiber doublets
(half at stereo angles of $\pm 3^{\circ}$) mounted on carbon
fiber tubes. 
Single hit $r$-$\phi$ resolutions are 80--100~$\mu$m.
Clear optical fibers carry the light to 77k channels of
visible light photon detectors and also provide a fast pick-off
for a track trigger.  Following a 2~T superconducting solenoid, 
a central (and forward at larger $\eta$'s) preshower detector of 
triangular scintillator strips
with embedded wavelength-shifting fibers improves non-isolated
electron identification to aid in semileptonic $b$ tagging.

Although the rate for $b$ production is high, the hadronic environment 
still results in a challenging signal to background, with a total
hadronic background of $\sigma_{\rm had}^{\rm tot} \approx 75$~mb to
be compared to $\sigma_{bb} \approx 0.1$~mb.  To address this, we have
a pipelined Level-3 trigger that will have the capability to trigger on tracks
with significant impact parameter at Level 2 and on tracks in given
momenta ranges at Level 1.  In addition, there will often be one or 
more overlapping minimum bias events on the event of interest, with an 
expected average of 2.0 at Run 2a's projected peak luminosity.

The projected performance of the \D0 upgrade detector is
a momentum resolution of $\delta p_T / p^2_T = 0.002$ combining the SMT and CFT
measurements and tracking out to $|\eta| < 3$ using the forward silicon disks.
The fitted primary vertex should have a resolution of 15--30~$\mu$m in 
$r$-$\phi$ and secondary vertices found with resolutions of 40~$\mu$m 
($r$-$\phi$) and 80~$\mu$m ($r$-$z$).  The upgrade detector has 
excellent lepton coverage for both triggers and identification:
muons in the range $p_T > 2.0$~GeV, $|\eta| < 2.0$ and 
electrons over $p_T > 2.0$~GeV, $|\eta| < 2.5$.  The impact parameter
resolution is expected to asymptotically approach 15~$\mu$m for
high-momentum tracks.
Comparisons of detector performance to these expected benchmarks will be
shown in the next section.


\section{Current Data and Performance}

The \D0 upgrade detector physically rolled in to place Jan.~2001 and
the first Run~2 collisions occurred in April 2001. Until Nov.~2001, 
activities were dominated by commissioning the silicon detector, 
establishing timing, and commissioning the DAQ and online systems.
Given the importance of tracking in $B$ physics, critical path 
items were late Analog Front End (AFE) boards that were essential for
reading out the central fiber tracker and preshower 
detectors.  In Summer 2001, only a very restrictive slice in $\phi$ 
was instrumented for CFT axial readout.  During a shutdown 
in Nov. 2001, a large fraction of CFT axial AFE boards were installed
and commissioned over the winter.  Due to this missing readout, many
commissioning tracking studies were performed with silicon-only tracking.
The silicon SMT detector itself is operating very well with 95\%
of the barrel sensors, 96\% of the small-$z$ F-disks, and 
87\% of the larger-$z$ H-disks operational.

Only by the end of winter 2002 were the CFT axial channels 
fully instrumented
and CFT stereo fully instrumented by the end of April 2002, i.e., it is 
only until very recently that \D0 has had its full tracking system 
available.  
Until May 2002, the Tevatron delivered approximately 35~pb$^{-1}$ and
the \D0 detector recorded about 10~pb$^{-1}$ of this in physics runs.

\begin{figure}[htb]
\begin{center}
\epsfig{file=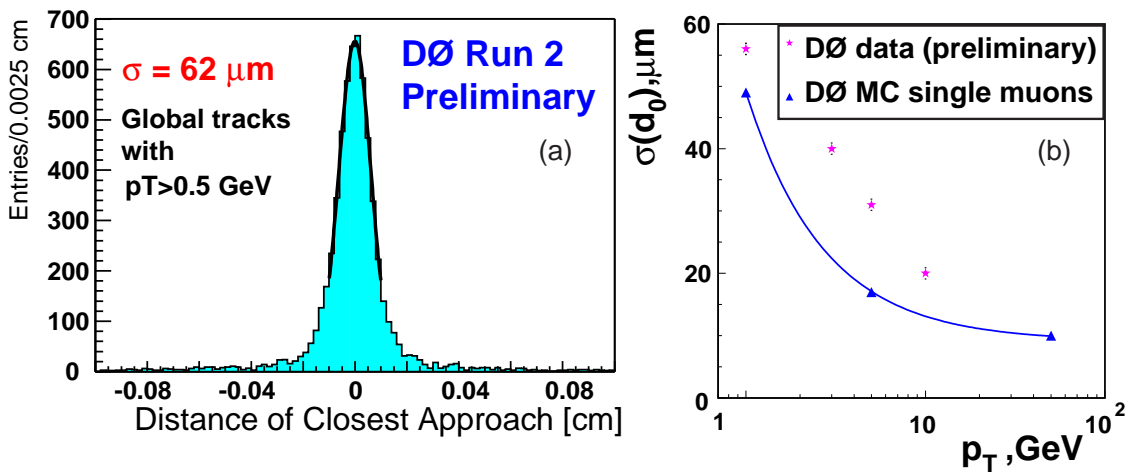,height=2.5in}
\caption{(a) Impact parameter resolution for \D0 global tracks
for $p_T > 0.5$~GeV and (b) behavior as a function of 
track $p_T$ compared to Monte Carlo expectations.}
\label{fig:impact}
\end{center}
\end{figure}

Paramount to being able to carry out the $B$ physics program is 
good impact parameter resolution and the ability to form precise
primary and secondary vertices.  Using current data, comparisons
between expectations and current performance are made.
For high multiplicity primary 
vertices (e.g., $N_{\rm track} \ge 14$), a primary vertex resolution 
of 46~$\mu$m
has been measured in the transverse $x$-$y$ plane.  Other 
measurements indicate this resolution contains a convolution
over a transverse beam size
of approximately 30~$\mu$m.
The impact parameter resolution measured with respect to
this primary vertex is currently found to be 62~$\mu$m for
global tracks (SMT and CFT point measurements) 
with transverse momentum $p_T > 0.5$~GeV as shown in
Fig.~\ref{fig:impact}(a). 
This asymptotically approaches a current value of 20~$\mu$m
for high-$p_T$ tracks and is compared to Monte Carlo expectations
in Fig.~\ref{fig:impact}(b).  It is clear that progress on alignment
of the tracking chambers has resulted in resolutions already approaching
expected values.

\section{Evidence for $b$ production}

\begin{figure}[htb]
\begin{center}
\epsfig{file=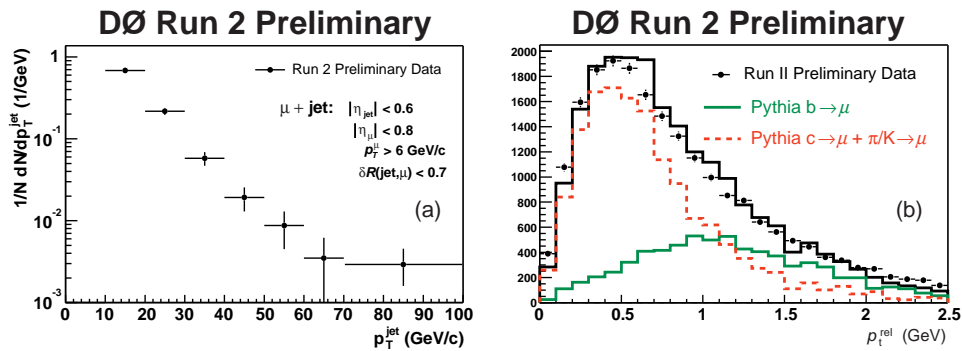,width=5.0in}
\caption{(a) Differential cross section for the rate of muons
associated with jets (see text) in the range $|\eta_{\rm jet}| < 0.6$,
$|\eta_{\mu}| < 0.8$, and $p_T^{\mu} > 6$~GeV; (b) fit to
the transverse momentum of the muon with respect to the muon plus
jet axis, i.e., $p_t^{\rm rel}$, to extract the $b$ rate.}
\label{fig:mujet}
\end{center}
\end{figure}

One of the first systems to be reliably commissioned was the muon system
and a benchmark analysis is the check of the muon plus jet rate
due to $b \rightarrow \mu$.  Muons reconstructed in only the
muon system without a central track match (due to commissioning delays
of the CFT) were associated with jets if close in angle:
$\delta R = \sqrt{\delta\phi^2 + \delta\eta^2} < 0.7$.  The 
measured differential
cross section is shown in Fig.~\ref{fig:mujet}(a) for the indicated
kinematic region.  The shape is consistent with MC predictions as
well as Run 1 \D0 results~\cite{run1mujet}.  Since the conference, the absolute
rate has been measured as well and confirmed to be consistent with
expectations.

The muons in this sample will be due to $b \rightarrow \mu$, 
$b \rightarrow c \rightarrow \mu$, $c \rightarrow \mu$, and
$\pi/K \rightarrow \mu$ decays in flight.  The decay muon from
the heavier $b$ quark will tend to get a larger transverse 
momentum kick
relative to the jet axis, $p_t^{\rm rel}$. Fig.~\ref{fig:mujet}(b)
shows typical fits in this parameter to extract the $b$ content.
Entries at large values of $p_t^{\rm rel}$ are due to $b$ hadrons,
and also provide a method for identifying $b$ jets.  Current work
using muons with a central track match shows substantial improvement
in separation of $b \rightarrow \mu$ from backgrounds in the same
parameter.

\begin{figure}[htb]
\begin{center}
\epsfig{file=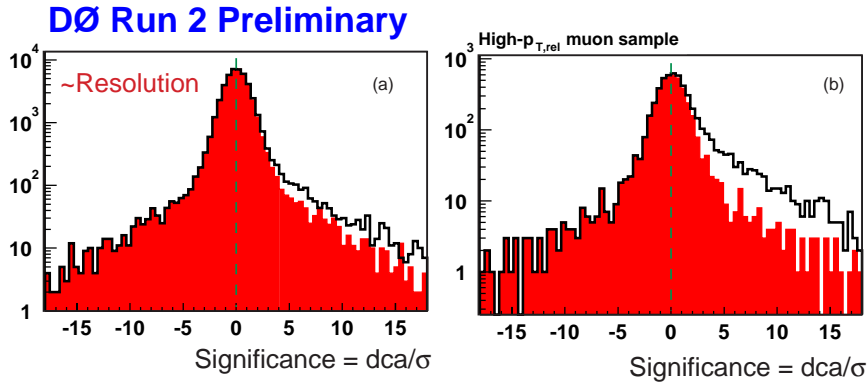,width=4.5in}
\caption{(a) Signed impact parameter significance 
(distance of closest approach (dca)
divided by its error) for tracks in an unbiased
di-jet sample with track $p_T > 1.5$~GeV, more than
10 total hits (SMT+CFT) and $|{\rm dca}| < 1.0$~mm.
The negative side of the distribution is also reflected
to the positive side. (b) The same distribution
in the muon plus jet sample with $p_t^{\rm rel} > 1.5$~GeV
to enhance the $b$-jet content.}
\label{fig:dcasig}
\end{center}
\end{figure}

Using this muon plus jet sample and demanding 
$p_t^{\rm rel} > 1.5$~GeV
to enhance the $b$ content, tracks can be examined for 
evidence of $b$ lifetime information.  A signed impact parameter
significance is formed where the sign is determined if the track
in question crosses the jet axis upstream (positive) or 
downstream of the found primary vertex.  
Figure~\ref{fig:dcasig}(a) shows the signed
impact parameter significance (i.e., the distance of 
closest approach (dca) divided by its error) for 
an unbiased di-jet sample for $|{\rm dca}| < 1.0$~mm to
reduce contributions due to $K^0_S$ and $\Lambda$.
The negative side of the distribution should be indicative
of the resolution, and its mirror image is also superimposed
on to the positive side of the distribution. The small excess is
due to residual $K^0_S$ and $\Lambda$ decays as well as small
amounts of $b$ jets in the sample.  Figure~\ref{fig:dcasig} shows
the same for the $b$-enhanced sample that shows a 
similar resolution as the di-jet sample, but a clear
excess of tracks with significant impact parameters due to
$b$ content.

\begin{figure}[htb]
\begin{center}
\epsfig{file=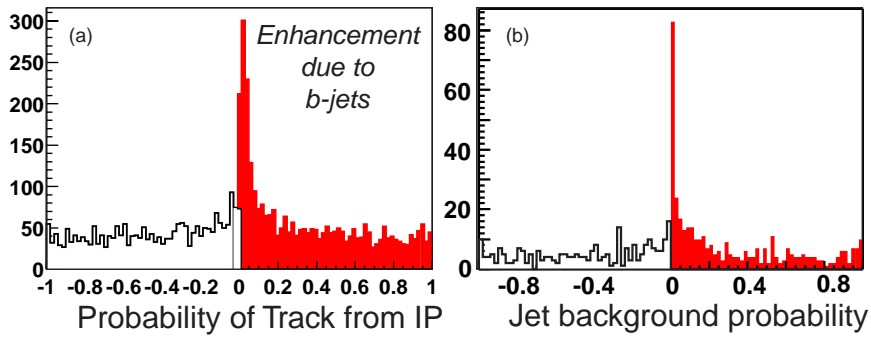,width=4.5in}
\caption{(a) From track significance 
values in the $p_t^{\rm rel} > 1.5$~GeV
muon plus jet sample, probability that a track is from the interaction
point or primary vertex; (b) normalized product of these track probabilities
for the tracks in the jet.  The peak is due to enhancement of $b$-jet content.}
\label{fig:impactprob}
\end{center}
\end{figure}

Using the di-jet sample, probability density functions can 
be formed for the observed resolution and Fig.~\ref{fig:impactprob}(a)
shows the resulting probability that tracks came from the
interaction point (IP) or primary vertex.  The spike on the positive
side near zero shows that there is an excess of tracks with 
small probability of coming from the IP, i.e., tracks from
secondary vertices carrying lifetime information. 
Taking the product of these track probabilities over the tracks
in a jet, the probability that the jet is due to a light quark
is shown in Fig.~\ref{fig:impactprob}(b).  This is the first
time that lifetime information has been measured in the \D0
detector.

\begin{figure}[htb]
\begin{center}
\epsfig{file=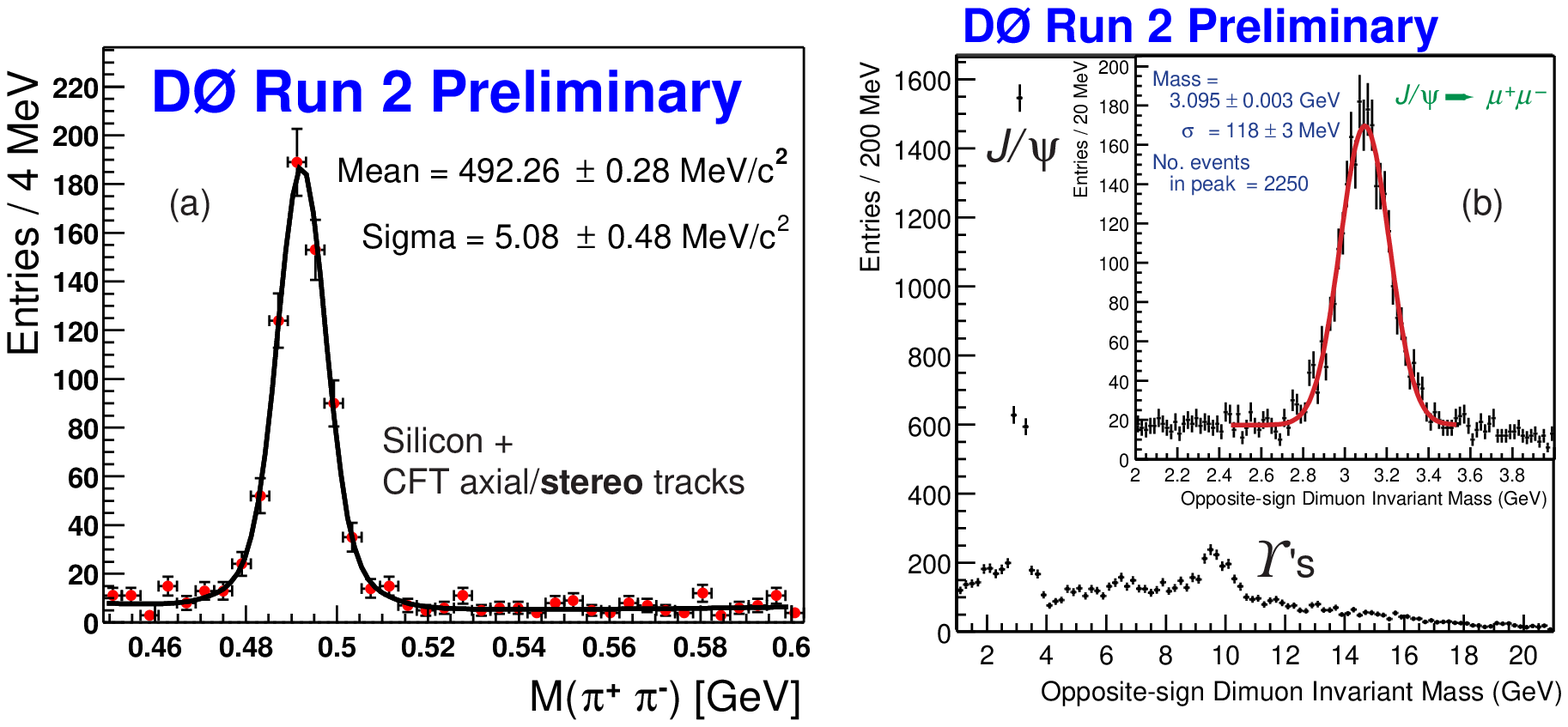,width=6.0in}
\caption{(a) Reconstructed $K^0_S \rightarrow \pi^+ \pi^-$ decays
for tracks with both CFT axial and stereo tracks; 
(b) reconstruction of $J/\psi \rightarrow \mu^+ \mu^-$ decays for
muons with a CFT central track match.}
\label{fig:masspeak}
\end{center}
\end{figure}

The next step in utilizing lifetime information is the formation of 
secondary vertices.  The important ``golden" channel 
of $B^0 \rightarrow J/\psi K^0_S$ to measure the 
$\sin{2\beta}$ parameter of CP violation serves as a
useful benchmark to evaluate detector performance.
The first physics objects reconstructed using
both CFT axial and stereo tracks are two-prong secondary
vertices from $K^0_S$ decays.  The invariant mass as shown in Fig.~\ref{fig:masspeak}(a) is found with a resolution of 
5.1~MeV, close to MC expectations of 5.0~MeV demonstrating that
the momentum resolution of the trackers is approaching
nominal values. 

\begin{figure}[htb]
\begin{center}
\epsfig{file=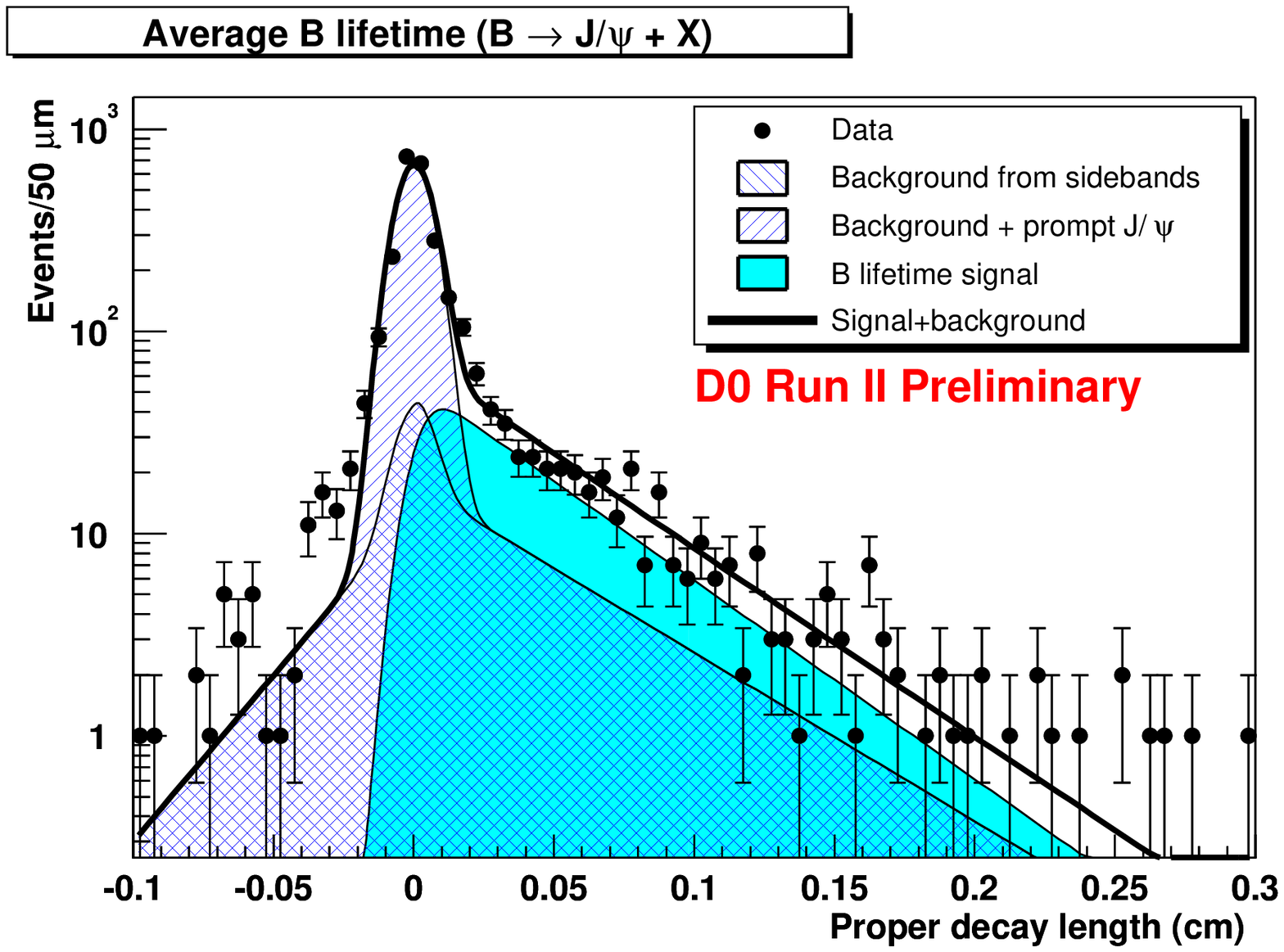,height=2.5in}
\caption{Proper decay length distribution for reconstructed
$J/\psi$ secondary vertices. The background is determined from
$J/\psi$ mass sidebands and prompt $J/\psi$ Monte Carlo.}
\label{fig:blife}
\end{center}
\end{figure}

Figure~\ref{fig:masspeak}(b) shows the
reconstruction of $J/\psi \rightarrow \mu^+\mu^-$ (and $\Upsilon$'s)
where the muons have a CFT track match.  The resultant mass
resolution of 118~MeV improves to approximately 70~MeV if
only tracks with both SMT and CFT hits are used.  This can be
compared to a resolution of 50--60~MeV expected from MC simulations.
After the conference, decay lengths of reconstructed $J/\psi$ secondary 
vertices due to $B \rightarrow J/\psi X$ as shown in Fig.~\ref{fig:blife}
have been used to find a lifetime consistent with the PDG~\cite{pdg}
value.

\section{Some expectations for Run 2a}

Continuing with expectations on the precision with which $\sin{2\beta}$ could
be measured using $B^0 \rightarrow J/\psi K^0_S$ events, the strengths of
the \D0 detector for such a $B$ physics measurement are summarized in 
Table~\ref{tab:sin2b}. One would measure the decay length
of the $B^0$ decaying into $J/\psi K^0_S$, and tag the $b$ quark flavor
at production using a same-side tag of pion charge (from the primary) or
an opposite-side tag by examining the lepton and/or jet charge.
Defining tagging efficiency as $\epsilon = N_{\rm tag}/N_{\rm tot}$, and the
dilution factor $D$ as the asymmetry between ``right" (R) and
``wrong" (W) tags, $D = (N_R - N_W)/(N_R + N_W)$, the overall flavor tag
quality can be quantified by $\epsilon D^2$.  Although the \D0 upgrade 
detector does not have kaon identification for opposite-side tagging, this
is balanced by excellent muon and electron coverage for lepton tagging, and
good forward tracking for determining jet charge.  We also expect to
be able to trigger on $J/\psi \rightarrow e^+e^-$ assisted by information 
from the preshower subdetectors.

\begin{table}[htb]
\begin{center}
\begin{tabular}{c|c|c}
\hline
Tag & \D0 Strength & Flavor Tag Quality, \\ 
    &              &  $\epsilon D^2$     \\    \hline
Same side tag    & --                                    & 2.0 \\
Soft lepton tag  & $\mu$ and $e$ coverage and identification & 3.1 \\
Jet charge tag   & forward tracking                          & 4.7 \\
Opposite-side kaon tag & {\it no $K$ identification}         &  -- \\ \hline
Combined         &                                           & 9.8 \\ \hline 
\end{tabular}
\caption{Summary of predicted performance and strengths of flavor
tagging in the CP violation channel $B^0 \rightarrow J/\psi K^0_S$.}
\label{tab:sin2b}
\end{center}
\end{table}

Using these projections, and the expectation of 30--40k reconstructed events
from the 2~fb$^{-1}$ of Run 2a, an error of $\delta \sin{2\beta} \approx 0.04$ 
is predicted~\cite{run2bphys}.

Another high priority $B$ physics analysis is the measurement of 
$B^0_s$ mixing.  Current limits exclude $x_s = \Delta m_s / \Gamma_s > 21$ 
at 95\% C.L.~\cite{pdg}, 
while a global fit to a number of CKM triangle measurements
in the framework of the Standard Model predicts~\cite{Ciuchini:2001zf}
$x_s = 25.2^{+2.2}_{-1.0}$, although new physics can easily result in a
much larger prediction.
This very important measurement should therefore be in reach of the 
Tevatron experiments.

\begin{figure}[htb]
\begin{center}
\epsfig{file=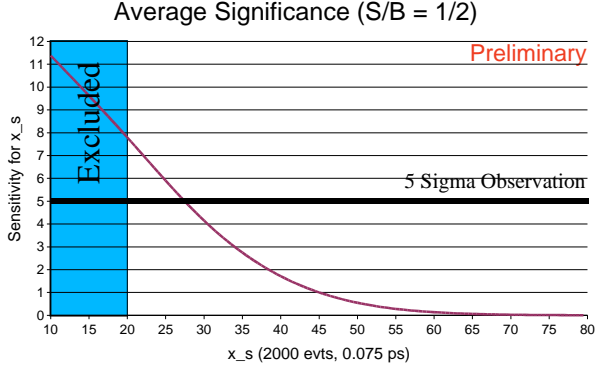,height=2.0in}
\caption{Projected reach on measuring $x_s$ in Run 2a in the hadronic
channel described in the text assuming a signal-to-background ratio of
0.5 and a proper decay time resolution of 0.098~ps.}
\label{fig:bsmixing}
\end{center}
\end{figure}

\D0 has made Monte Carlo studies of both the semileptonic and hadronic
decays of $B^0_s$ to extract this mixing.  In the hadronic mode, which has
the advantage of no missing neutrino, 
decays of $B^0_s \rightarrow D^-_s \pi^+ (\pi^+ \pi^-)$; 
$D_s^- \rightarrow \phi \pi^-$, and $\phi \rightarrow K^+ K^-$ can
be reconstructed.  Triggers will be from leptons on the opposite side, and
this lepton charge will tag the initial flavor. The final flavor will
be tagged by the charge of the $D_s$.  Approximately 2000 events are
expected in 2~fb$^{-1}$, and with a signal-to-background ration of 0.5, 
and a pessimistic proper time resolution of 0.098~ps, the resulting
projected reach on measuring $x_s$ for this channel is shown 
in Fig.~\ref{fig:bsmixing}.

\section{Future}

The \D0 upgrade detector with its tracking upgrades is much better suited
for $B$ physics than the Run 1 detector.  The detector is still being 
commissioned, but we have observed lifetime information from $b$ hadrons
for the first time in the \D0 detector,
and are approaching data quality to soon allow preliminary $B$ physics 
results.

Detector commissioning will continue with debugging, calibration, and 
alignment of the subdetectors and refinement of reconstruction algorithms 
to select physics objects.  The full tracking system has just recently
been available and secondary vertexing is rapidly improving as well 
as the prospects for tagging electrons (for $b$ quark identification
and $J/\psi \rightarrow e^+e^-$) using a road method and the preshower
subdetectors.

The next large jump in performance will come from increasing the scope
of the trigger system.  The Level 2 trigger is coming online, a Level 1 
central track trigger is expected at the end of summer 2002, and a
Level 2 silicon track trigger able to fire on tracks with large impact
parameter significance should be available fall 2002.  

Of course, what 
is also needed is integrated luminosity and the Tevatron's performance
has fallen short of expectations.  Task forces at Fermilab have been
addressing this low luminosity, and integrated luminosities of 
300~pb$^{-1}$ by the end of 2002 have been promised by laboratory 
management.  The \D0 collaboration eagerly looks forward to this
data and the $B$ physics that it holds for the future.

\end{document}